\documentclass[aps,11pt]{revtex4}
\usepackage[dvips]{graphicx}

\usepackage{amsmath}
\usepackage{bm}

\begin{document}

\preprint{}
\title{Unification versus proton decay in $SU(5)$}
\author{Ilja Dor\v{s}ner}
\email{idorsner@ictp.it} \affiliation{The Abdus Salam
International Centre for Theoretical Physics\\
Strada Costiera 11, 34014 Trieste, Italy}
\homepage{http://www.ictp.it/~idorsner}
\author{Pavel Fileviez P\'erez}
\email{fileviez@cftp.ist.utl.pt} \affiliation{Centro de
F{\'\i}sica Te\'orica de Part{\'\i}culas.\ Departamento de
F{\'\i}sica.\ Instituto Superior T\'ecnico.\ Avenida Rovisco Pais,
1\\ 1049-001 Lisboa, Portugal}
\homepage{http://cftp.ist.utl.pt/~fileviez}
\begin{abstract}
We investigate unification constraints in the simplest
renormalizable non-supersymmetric $SU(5)$ framework. We show that
in the scenario where the Higgs sector is composed of the
$\bm{5}$, $\bm{24}$, and $\bm{45}$ dimensional representations the
proton could be practically stable. We accordingly demonstrate
that of all the $SU(5)$ scenarios only the non-renormalizable one
with the $\bm{5}$, $\bm{24}$, and $\bm{15}$ dimensional Higgs
multiplets can be verify if low-energy supersymmetry is not
realized in Nature.
\end{abstract}
\maketitle
\section{Introduction}
Grand unified theories (GUTs) are considered to be among the most
appealing scenarios for physics beyond the Standard Model.
Qualitatively they always predict (i) unification of gauge
couplings of the Standard Model and (ii) proton decay. The first
feature cannot be directly probed since unification takes place at
a very high energy scale---the so-called GUT scale. However the
second feature \emph{can} be probed and it offers the only
realistic way of testing grand unification. It is thus important
to single out and investigate viable models of grand unification;
where proton decay is not only well predicted, but also
experimentally accessible in both current and future proton decay
experiments.

Out of all grand unified theories the scenarios based on $SU(5)$
gauge symmetry are arguably the most predictive ones for proton
decay. Recall, of all simple gauge groups that allow SM embedding
only $SU(5)$ has unique single-step symmetry breaking pattern.
This allows rather accurate determination of high energy scales
relevant for proton decay. And, in its non-supersymmetric version,
$SU(5)$ GUT avoids uncertainties pertaining to the exact nature as
well as the relevant scale of supersymmetry breaking; both of
those features are yet to be established experimentally. Moreover,
it simplifies the discussion on the dominant source(s) of proton
decay. All these appealing properties single out
non-supersymmetric $SU(5)$ as the theory for proton decay. We
accordingly focus our attention on its simplest realistic
realizations.

Our starting point is the $SU(5)$ model proposed long ago by
H.~Georgi and S.~Glashow~\cite{GG}. Their model offers partial
matter unification by accommodating $i$th generation of matter
fields in the $\overline{\bm{5}}_i$ and $\bm{10}_i$ dimensional
representations. The scalar sector is composed of a $\bm{24}$
dimensional Higgs representation and a $\bm{5}$ dimensional Higgs
multiplet. The SM singlet in $\bm{24}$ breaks $SU(5)$ symmetry
down to the Standard Model, while the SM $SU(2)$ doublet in
$\bm{5}$ accomplishes electroweak symmetry breaking. The model,
however, is not realistic; the gauge couplings do not unify,
neutrinos are massless and $m_{\mu(e)}=m_{s(d)}$ at the GUT scale.

There are two possible model building approaches that lead to
simple yet realistic extensions of the Georgi-Glashow (GG) model
in view of generation of realistic charged fermion masses. One
approach is to allow for higher-dimensional operators which modify
bad mass relations $m_{\mu(e)}=m_{s(d)}$~\cite{Ellis:1979fg}. That
approach requires no additional Higgs fields to be introduced to
fix those relations. And, the strength of required corrections in
the charged sector might allow one to place an upper bound on the
scale where the UV completion of the unified theory takes place.
Of course, in order to improve unification more split
representations need to be present. The other approach is to stick
with renormalizable operators. The latter approach requires
addition of a $\bm{45}$ dimensional Higgs multiplet. (Recall that
the tensor product $\bm{10} \otimes \overline{\bm{5}} = {\bm{5}}
\oplus \bm{45}$.) Within that approach, as far as the neutrino
masses are concerned, one can either introduce some fermion
singlets---the right-handed neutrinos---or a $\bm{15}$ dimensional
Higgs representations (or both).

The simplest non-renormalizable model based on $SU(5)$ has been
subject of a recent investigation~\cite{Dorsner:2005fq}. It has been
shown that the model, with the Higgs sector composed of $\bm{5}$,
$\bm{24}$ and $\bm{15}$, can be tested at future proton decay
experiments and at future collider experiments, for example at
LHC. The first possibility is due to existence of an upper bound
on the proton decay lifetime. Namely, $\tau_p \leq 1.4 \times
10^{36} (0.015\,\textrm{GeV}^3/\alpha)^2$\,years, where $\alpha$
is the nucleon matrix element. The second one is based on
potential production of light leptoquarks~\cite{Dorsner:2005fq}.
See reference~\cite{Dorsner:2005ii} for the study of several
phenomenological and cosmological issues in this context.

In this work we want to investigate the minimal extension of the
Georgi-Glashow model within the renormalizable framework. Namely,
we want to study predictions of a model with the Higgs sector made
out of $\bm{5}$, $\bm{45}$ and $\bm{24}$ representations. We
initially assume that the matter sector contains right-handed
neutrinos to generate neutrino masses through the Type-I see-saw
mechanism~\cite{seesaw}. However, we also discuss the case when
there is an extra $\bm{15}$ dimensional representation that
generates neutrino mass via Type-II see-saw~\cite{seesaw2}.

The paper is organized as follows: In Section I we describe the
minimal renormalizable $SU(5)$ and consequently study the
unification constraints. In Section II we compare different
scenarios based on $SU(5)$ and discuss possibility to test them.
In the last section we summaries our results.

\section{Minimal renormalizable $SU(5)$}

The Higgs sector of the minimal renormalizable $SU(5)$ is composed
of the $\bm{5}$, $\bm{24}$, and $\bm{45}$ dimensional
representations. Their SM $SU(3) \times SU(2) \times U(1)$
decomposition is given by:
\begin{eqnarray*}
\bm{5}&=& H_1 + T = (\bm{1},\bm{2},1/2)+(\bm{3},\bm{1},-1/3),\\
\bm{24}&=&\Sigma_8 + \Sigma_3 + \Sigma_{(3,2)} +
\Sigma_{(\overline{3},2)} + \Sigma_{24}\\
&=& (\bm{8},\bm{1},0)+(\bm{1},\bm{3},0)
+(\bm{3},\bm{2},-5/6)+(\overline{\bm{3}},\bm{2},5/6)
+(\bm{1},\bm{1},0),\\
\bm{45}&=&\Phi_1 + \Phi_2 + \Phi_3 + \Phi_4 +\Phi_5 + \Phi_6 +
H_2\\
&=& (\bm{8},\bm{2},1/2)+ (\overline{\bm{6}},\bm{1}, -1/3) +
(\bm{3},\bm{3},-1/3) + (\overline{\bm{3}}, \bm{2}, -7/6) \\
&+& (\bm{3},\bm{1}, -1/3) + (\overline{\bm{3}}, \bm{1}, 4/3) +
(\bm{1}, \bm{2}, 1/2), \nonumber
\end{eqnarray*}
where we also set our notation. The Yukawa potential for charged
fermions read as:
\begin{eqnarray}
V_{Yukawa} &=& (Y_1)_{ij} \ \bm{10}_i^{\alpha \beta} \
\overline{\bm{5}}_j^{\alpha} \ (\bm{5}^*_H)^{\beta} \ + \ (Y_2)_{ij}
\ \bm{10}_i^{\alpha \beta} \ \overline{\bm{5}}^{\delta}_j
\ (\bm{45}^*_H)^{\alpha \beta}_{\delta} \ + \nonumber\\
&+&\epsilon_{\alpha \beta \gamma \delta r} \left( (Y_3)_{ij} \
\bm{10}_i^{\alpha \beta} \ \bm{10}_j^{\gamma \delta} \ \bm{5_H}^r \
+ \ (Y_4)_{ij} \ \bm{10}_i^{\alpha \beta} \bm{10}_j^{m \gamma}
(\bm{45_H})^{\delta r}_m \right),\qquad i=1,..,3,
\end{eqnarray}
where the field $\bm{45}$ satisfies the following conditions:
\begin{eqnarray}
(\bm{45})^{\alpha \beta}_{\delta} &=& - (\bm{45})^{\beta
\alpha}_{\delta}, \ \
\sum_{\alpha=1}^5 (\bm{45})^{\alpha \beta}_{\alpha} = 0 \\
\sum_{i=1}^3 <\bm{45}>^{i 5}_{i} &=& - <\bm{45}
>^{45}_4\qquad(v_{45} = <\bm{45}>^{1 5}_{1}=
<\bm{45}>^{2 5}_{2}=<\bm{45}>^{3 5}_{3}).
\end{eqnarray}

In this model the masses for charged fermions are given by:
\begin{eqnarray}
M_D &=& Y_1 \ v_5^* \ + \ 2 \ Y_2 \ v_{45}^* ,\\
M_E &=& Y^T_1 \ v_5^* \ - 6 \ Y^T_2 \ v_{45}^* , \label{GJ}\\
M_U &=& 4 \ (Y_3+Y^T_3) \ v_5 \ -  \ 8 \ (Y_4-Y^T_4) \ v_{45},
\end{eqnarray}
where $<\bm{5}>=v_5$. $Y_1$, $Y_2$, $Y_3$ and $Y_4$ are arbitrary
$3 \times 3$ matrices. (Note the Georgi-Jarlskog~\cite{GJ} factor
in Eq.~(\ref{GJ}).) Clearly, there are enough parameters in the
Yukawa sector to fit all charged fermions masses. For previous
studies in this context see~\cite{45studies}. Now, let us
understand the unification constraints within this model.

\subsection{Unification of gauge interactions}

Necessary conditions for the successful gauge coupling unification
can be expressed via two equalities. (See reference~\cite{Giveon}
for details.) These are
\begin{equation}
\frac{B_{23}}{B_{12}}=\frac{5}{8} \frac{\sin^2
\theta_w-\alpha_{em}/\alpha_s}{3/8-\sin^2 \theta_w},\qquad\qquad
\ln \frac{M_{GUT}}{M_Z}=\frac{16 \pi}{5 \alpha_{em}}
\frac{3/8-\sin^2 \theta_w}{B_{12}}.
\end{equation}
where all experimentally measured quantities on the right-hand
sides are to be taken at $M_Z$ energy scale. The first one is the
so-called ``B-test'' and the second one is the ``GUT scale
relation''. In what follows we use~\cite{PDG} $\sin^2
\theta_w=0.23120 \pm 0.00015$, $\alpha_{em}^{-1}=127.906 \pm
0.019$ and $\alpha_{s}=0.1187 \pm 0.002$ to obtain:
\begin{equation}
\label{condition1} \frac{B_{23}}{B_{12}}=0.719\pm0.005,\qquad
\qquad\ln \frac{M_{GUT}}{M_Z}=\frac{184.9 \pm 0.2}{B_{12}}.
\end{equation}
The left-hand sides, on the other hand, depend on particular
particle content of the theory at hand and corresponding mass
spectrum. More precisely, $B_{ij}=B_i - B_j$, where $B_i$
coefficients are given by:
\begin{equation}
\label{r} B_i = b_i+\sum_{I} b_{iI} r_{I}, \qquad r_I=\frac{\ln
M_{GUT}/M_{I}}{\ln M_{GUT}/M_{Z}}.
\end{equation}
$b_i$ are the SM coefficients while $b_{iI}$ are the one-loop
coefficients of any additional particle $I$ of mass $M_I$ ($M_Z
\leq M_I \leq M_{GUT}$). (Recall, for the case of $n$ light Higgs
doublet fields $b_1=40/10+n/10$, $b_2=-20/6+n/6$ and $b_3=-7$.)
Relevant $B_{ij}$-coefficient contributions in our scenario are
listed in Table~\ref{table1}.
\begin{table}[h]
\caption{\label{table1} Contributions to the $B_{ij}$
coefficients. The masses of the Higgs doublets are taken to be at
$M_Z$.}
\begin{ruledtabular}
\begin{tabular}{lccccccccccccc}
 & 2HSM & $T$ & $V$ & $\Sigma_8$ & $\Sigma_3$ & $\Phi_1$ & $\Phi_2$ & $\Phi_3$ & $\Phi_4$ & $\Phi_5$ & $\Phi_6$\\
\hline $B_{23}$ & $4$ & $-\frac{1}{6} r_{T}$ & $-\frac{7}{2} r_V$
& $-\frac{1}{2} r_{\Sigma_8}$ & $\frac{1}{3} r_{\Sigma_3}$ &
$-\frac{2}{3}r_{\Phi_1}$ & $- \frac{5}{6} r_{\Phi_2}$ &
$\frac{3}{2} r_{\Phi_3}$ & $\frac{1}{6} r_{\Phi_4}$
& $-\frac{1}{6} r_{\Phi_5}$ &  $-\frac{1}{6} r_{\Phi_6}$\\
$B_{12}$ & $\frac{36}{5}$ & $\frac{1}{15} r_{T}$ & $-7 r_V$ & 0 &
$-\frac{1}{3} r_{\Sigma_3}$ & $-\frac{8}{15} r_{\Phi_1}$
&$\frac{2}{15} r_{\Phi_2}$ & $-\frac{9}{5} r_{\Phi_3}$
& $\frac{17}{15} r_{\Phi_4}$ & $\frac{1}{15} r_{\Phi_5}$ & $\frac{16}{15} r_{\Phi_6}$\\
\end{tabular}
\end{ruledtabular}
\end{table}

There are five SM multiplets that mediate proton decay in this
model. These are the superheavy gauge bosons
$V(=(\bm{3},\bm{2},-5/6)+(\overline{\bm{3}},\bm{2},5/6))$, the
$SU(3)$ triplet $T$, $\Phi_3$, $\Phi_5$ and $\Phi_6$. The least
model dependent and usually the most dominant proton decay
contribution comes from gauge boson mediation. Its strength is set
by $M_V$ and $\alpha_{GUT}$---the value of gauge coupling at
$M_{GUT}$. In what follows, we identify $M_V$ with the GUT scale,
i.e., we set $M_V \equiv M_{GUT}$. Clearly, we are interested in
the regime where $M_V(=M_{GUT})$ is above experimentally
established bounds. Now, how large $M_{GUT}$ is primarily depends
on masses of $\Sigma_3$, $\Phi_1$, and $\Phi_3$ through their
negative contribution to $B_{12}$. If they are light enough they
render gauge contributions to proton decay innocuous. However,
$\Phi_3$ field cannot be very light due to proton decay
constraints. At the same time, it cannot be at the GUT scale since
B-test cannot be satisfied using solely $\Sigma_3$ and/or
$\Phi_1$. Clearly, proton decay constraints will thus create
tension between successful unification and possible values for
$M_{\Phi_3}$ and $M_{GUT}$.

We again note that contributions from fields in $\Sigma_3$ and
$\Phi_1$ cannot sufficiently modify B-test. This is because the SM
fails rather badly, i.e., $B^{SM}_{23}/B^{SM}_{12}\simeq 0.51$, so
that large corrections to $B_{23}$ and $B_{12}$ are needed. Thus,
we always need to use contribution coming from the field $\Phi_3$
to some extent. This contradicts previous studies~\cite{BabuMa}
where successful unification was claimed with $\Phi_3$ field kept
at the GUT scale. Unification constraints in the context of the
model with the same Higgs content as ours have also been studied
before in~\cite{Giveon}. However, authors did not notice that
$\Phi_3$ in general mediates nucleon decay. Moreover, even if the
model violates baryon number single $\bm{45}$ dimensional
representation is sufficient for successful unification contrary
to the remarks in Ref.~\cite{Giveon}.

The $\Phi_3$ contributions to proton decay are coming from
interactions $Y_4 Q^T \imath \sigma_2 \Phi_3 Q$ and $Y_2 Q^T
\imath \sigma_2 \Phi_3^* L$ (for a review on proton decay
see~\cite{ProtonReview}). Our calculation shows that $\Phi_3$
should be heavier than $10^{10}$\,GeV in order not to conflict
experimental data. (Of course, this rather naive estimate holds if
one assumes most natural values for Yukawa couplings.) If for some
reason one of the two couplings is absent or suppressed the bound
on $\Phi_3$ would seize to exist. For example, if we choose $Y_4$
to be antisymmetric matrix, the coupling $Y_4 Q^T \imath \sigma_2
\Phi_3 Q$ vanishes. Therefore, $\Phi_3$ could be very light.

\begin{figure}[h]
\begin{center}
\includegraphics[width=5in]{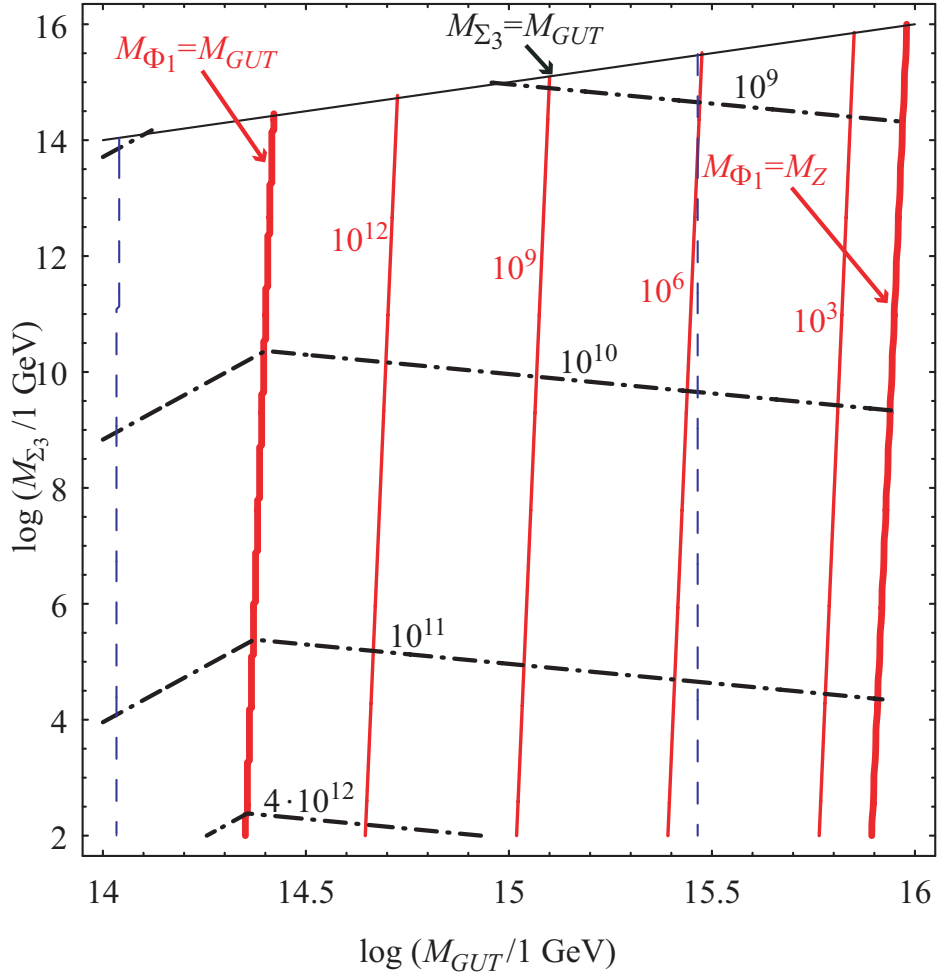}
\end{center}
\caption{\label{figure:1} Plot of lines of constant value of
$M_{\Phi_1}$ (solid line) and $M_{\Phi_3}$ (dash-dot line) in the
$\textit{log} (M_{GUT}/1\,\textrm{GeV})$--$\textit{log}
(M_{\Sigma_3}/1\,\textrm{GeV})$ plane. To generate the plot we
require exact one-loop unification with central values for the
gauge couplings as given in the text. All the masses are in the
GeV units. The viable gauge coupling unification region is bounded
from the right (below) by the requirement that $M_{\Phi_1}
(M_{\Sigma_3}) \geq M_Z$. Two dashed lines represent lower bounds
on the GUT scale due to the proton decay experimental limits. The
left (right) one is generated under the assumption of
suppression (enhancement) of the flavor dependent part of proton
decay amplitudes. Both lines correspond to $\alpha=0.015$\,GeV$^3$.}
\end{figure}

There are four critical mass parameters ($M_{\Phi_1}$,
$M_{\Phi_3}$, $M_{\Sigma_3}$ and $M_{GUT}$) and two equations that
govern unification. So, we show in Fig.~\ref{figure:1} a contour
plot of $M_{\Phi_1}$ (solid line) and $M_{\Phi_3}$ (dash-dot line)
in the $M_{GUT}$--$M_{\Sigma_3}$ plane in order to present the
full parameter space for successful gauge coupling unification.

Fig.~\ref{figure:1} shows that $\Sigma_3$ alone cannot generate
unification. So, the proton decay mediating field $\Phi_3$ needs
to be below the GUT scale. Its mass varies between $10^9$\,GeV and
$10^{12}$\,GeV in the shown region. Recall, $M_{\Phi_3}$ must be
above $ 10^{10}$\,GeV unless some additional symmetry or
cancellation is assumed. And, the lighter the $\Phi_3$ field is
the higher the GUT scale gets. Basically, change in the $\Phi_3$
mass by a factor of $10^3$ corresponds to change in the GUT scale
by a factor of 10. This can be traced back to its rather
significant impact on $B_{12}$.

The viable gauge coupling unification region in
Fig.~\ref{figure:1} is bounded from the right (below) by
requirement that $M_{\Phi_1} (M_{\Sigma_3}) \geq M_Z$. And, the
plot is valid only in the region where $M_{\Sigma_3} \leq
M_{GUT}$. Clearly, there are two qualitatively distinct regions
separated by the $M_{\Phi_1} = M_{GUT}$ curve. To the left of the
$M_{\Phi_1} = M_{GUT}$ curve only $\Sigma_3$ and $\Phi_3$ play the
role in unification and hence the change in slope of $M_{\Phi_3}$
as one crosses it.

To help the reader we also plot current bounds (dashed lines) on
the GUT scale that stem from experimental bounds on proton decay
lifetime and the $M_V=M_{GUT}$ relation. There are two of them.
The one on the left corresponds to lower bound on the GUT scale in
the case of suppression of the flavor dependent part of the total
proton decay amplitude~\cite{Dorsner:2005fq}. The right one
corresponds to maximally enhanced partial amplitude for $p
\rightarrow \pi^0 e^+$. (In both cases we use experimental limit
$\tau_p \geq 5.0 \times 10^{33}$\,years~\cite{PDG}.) More
precisely, by using the flavor freedom of the $d=6$ gauge mediated
proton decay amplitudes, one can specify lower bounds for
suppressed (enhanced) scenario on the GUT scale to be $M_{GUT}
\geq 3.0 \times 10^{14} \sqrt{\alpha_{GUT}}
\sqrt{\alpha/0.003\,\textrm{GeV}^3}$ ($M_{GUT} \geq 8.0 \times
10^{15} \sqrt{\alpha_{GUT}} \sqrt{\alpha/0.003\,\textrm{GeV}^3}$).
The lines shown are generated for
$\alpha=0.015$\,GeV$^3$~\cite{lattice}, where $\alpha$ is the
nucleon matrix element. Note that any ``intermediate'' scenario
for fermion masses falls in between in terms of the $M_{GUT}$
bounds. (See~\cite{DorsnerFileviez} for more details.) Clearly,
realistic scenario where only $\Phi_3$ corrects the SM running
with all other fields at $M_{GUT}$ is possible.

Fig.~\ref{figure:1} was generated under simplifying assumption
that only $\Sigma_3$, $\Phi_1$, and $\Phi_3$ are allowed to be
below $M_{GUT}$. But, in general, other fields could venture below
the GUT scale too. If we allow for such a scenario and place a
lower limit on $\Phi_3$ mass to be $10^{10}$\,GeV in order to
avoid rapid proton decay the maximal value of the GUT scale comes
out to be $3 \times 10^{16}$\,GeV. At the same time
$M_{\Sigma_3}=M_{\Phi_1}=M_Z$ and $M_{\Sigma_8}=4 \times
10^{5}\,\textrm{GeV}$. All other fields play no significant role
and are at the GUT scale.

In the previous discussion we have assumed that superheavy
right-handed neutrinos generate observed neutrino masses. If we do
not want singlets in the theory, we have to introduce the
$\bm{15}$ of Higgs to generate neutrino masses through the type II
see-saw mechanism. There is a difference between the two scenarios
from the point of view of proton decay. Namely, the singlets do
not significantly affect the running while split multiplets in
$\bm{15}$ could do that. Moreover, $\bm{15}$ contains scalar
leptoquarks that, through the mixing with the $\bm{5}$ of Higgs
could also mediate proton decay. (Recall, $\bm{15}= \Phi =(\Phi_a,
\Phi_b, \Phi_c)= (\bm{1},\bm{3},1)+
(\bm{3},\bm{2},1/6)+(\bm{6},\bm{1},-2/3)$.) In the case that
$\bm{15}$ is included in the model additional contributions to the
$B_{ij}$ are:
\begin{table}[h]
\caption{\label{tab:table1} Contributions of an extra $\bm{15}$ to
$B_{ij}$ coefficients.}
\begin{ruledtabular}
\begin{tabular}{lcccc}
     & $\Phi_a$ & $\Phi_b$ & $\Phi_c$\\
\hline $B_{23}$&$\frac{2}{3}r_{\Phi_a}$
&$\frac{1}{6} r_{\Phi_b}$ &$-\frac{5}{6} r_{\Phi_c}$\\
$B_{12}$&$-\frac{1}{15}r_{\Phi_a}$ &$-\frac{7}{15} r_{\Phi_b}$ &$\frac{8}{15} r_{\Phi_c}$\\
\end{tabular}
\end{ruledtabular}
\end{table}

There are two fields in $\bm{15}$ that can improve unification;
these are $\Phi_a$ and $\Phi_b$. $\Phi_a$ has very small
contribution to $B_{12}$ and very large contribution to $B_{23}$.
This means that its impact on the GUT scale is not significant.
$\Phi_b$, on the other hand, has large impact on the GUT scale
relation but, in general, it mediates proton decay and it is
probably better to keep it heavy.

In any case, the $\Phi_3$ contribution to the running of the gauge
couplings is crucial to achieve high scale unification in
agreement with experimental data. If its contribution to the decay
of the proton is set to zero by additional symmetry the
unification scale could be very large. Therefore, since in that
case the most important contributions to the decay of the proton
are the gauge $d=6$ ones, we can conclude that proton could be
\textit{stable}\/ for all practical purposes in the minimal
renormalizable $SU(5)$.

\subsection{Testing minimal realistic $SU(5)$ models}

As we discussed in previous sections there are two simple
candidates for unification based on $SU(5)$. In the first scenario
the Higgs sector is composed of $\bm{5}$, $\bm{24}$, and
$\bm{15}$, and higher-dimensional operators are used to modify the
relation $m_{\mu(e)}=m_{s(d)}$~\cite{Dorsner:2005fq}. Let us call
this model GUT-I. The second scenario is the one discussed in this
work---the renormalizable model with a Higgs sector composed of
$\bm{5}$, $\bm{24}$, and $\bm{45}$. Let us call this GUT-II. There
are two possibilities to test the GUT-I~\cite{Dorsner:2005fq}. One
is through proton decay in the current and next generation of
experiments and the other is through the production of light
leptoquarks in future colliders. In this work we have concluded
that GUT-II model cannot be tested through proton decay since the
lifetime of the nucleon could be very large. Therefore, we can say
that the only GUT candidate based on $SU(5)$ which can be
falsified in the near future is the model presented in
reference~\cite{Dorsner:2005fq}.

\section{Summary}

We have studied the possibility to achieve unification without
supersymmetry in a minimal realistic grand unified theory based on
$SU(5)$, where the Higgs sector is composed of the $\bm{5}$,
$\bm{24}$, and $\bm{45}$ dimensional representations. We have
pointed out that the proton could be practically stable in this
scenario. We have accordingly concluded that the best candidate to
be tested is the GUT model with $\bm{5}$, $\bm{24}$, and $\bm{15}$
dimensional representations in the Higgs sector.

\begin{acknowledgments}
{\small I.~D.~thanks Stephen M.~Barr and Bartol Research Institute
for support and hospitality. The work of P.~F.~P has been
supported by {\em Funda\c{c}\~{a}o para a Ci\^{e}ncia e a
Tecnologia} (FCT, Portugal) through the project CFTP,
POCTI-SFA-2-777 and a fellowship under project
POCTI/FNU/44409/2002. We would like to thank Goran Senjanovi\'c
for discussions and comments.}
\end{acknowledgments}


\end{document}